\documentclass[conference]{IEEEtran}
\usepackage{graphicx}
\graphicspath{ {images/} }
\usepackage{amssymb,amsthm,amsmath}

\usepackage{amssymb}

\ifCLASSINFOpdf
\else
\fi
\begin{document}
%
\title{Matrix Coherency Graph: A Tool for Improving 	Sparse Coding Performance}

\author{\IEEEauthorblockN{Mohsen Joneidi$^\dagger$, Mahdi Barzegar Khalilsarai$^\ast$, Alireza Zaeemzadeh$^\dagger$, Nazanin Rahnavard$^\dagger$}
\IEEEauthorblockA{
$^\dagger$University of Central Florida, Florida, US\\
$^\ast$Sharif University of Technology, Tehran, Iran\\
}
\and

}


%


\maketitle

\begin{abstract}
Exact recovery of a sparse solution for an underdetermined system of linear equations implies full search among all possible subsets of the dictionary, which is computationally intractable, while 
$l_1$
minimization will do the job when a Restricted Isometry Property holds for the dictionary. Yet, practical sparse recovery algorithms may fail to recover the vector of coefficients even when the dictionary deviates from the RIP only slightly. To enjoy
 $l_1$  minimization guarantees in a wider sense, a method based on a combination of full-search and 
$l_1$
 minimization is presented. The idea is based on partitioning the dictionary into atoms which are in some sense well-conditioned and those which are ill-conditioned. Inspired by that, a matrix coherency graph is introduced which is a tool extracted by the structure of the dictionary. This tool can be used for decreasing the greediness of sparse coding algorithms so that recovery will be more reliable. We have modified the IRLS algorithm by applying the proposed method on it and simulation results show that the modified version performs quite better than the original algorithm.\\
\\
\textit{Keywords-}
$l_1$minimization, Restricted Isometry Property, matrix coherency graph, IRLS
\end{abstract}


%
\IEEEpeerreviewmaketitle

\section{Introduction}
Consider an underdetermined system of linear equations:\\
\begin{equation}
\textbf{D \textit{x = y}}
\end{equation}
\\Where 
$\textbf{D} \in \mathbb{R}^{N\times K}$
is the dictionary containing proper bases, and 
$\textbf{\textit{y}} \in \mathbb{R}^{N}$ is an observation vector which is supposed to be a linear combination of a few columns of
$\textbf{D}$. Our goal is to recover the sparse solution
$\textbf{\textit{x}}\in \mathbb{R}^K$ which contains the sparse coefficients of the combination. The problem of recovery of 
\textbf{\textit{x}}
 can be written as
$(P_0)$:
\textbf{
\begin{equation}
\vec{\textbf{\textit{x}}}_{K \times 1}  = \underset{z\in  \mathbb{R}^K}{\arg\min} ~\Vert y-\textbf{D}z\Vert_2 ^2 ~~~~ s.t. ~~ \Vert z \Vert_0 \le s
\end{equation}
}\\
$\Vert . \Vert_0$ represents the number of non-zero elements of a vector and is actually a pseudo-norm. Exact solution of the above optimization problem is through the combinational search among all possible dictionary subsets. Due to its high computational burden, this algorithm is impractical for high dimension scenarios. Practical algorithms for sparse recovery include two major methods: greedy matching algorithms and, convex relaxation methods. The core idea of greedy methods is to involve proper basis vectors sequentially, where in each sequence the best atom in the sense of a criterion is selected to decrease the residual error of the previously selected bases. Orthogonal Matching Pursuit [1], Least Angle Regression, Forward Stage-wise Regression [2] and Subspace Pursuit [3] are examples of the first approach. On the other hand, relaxation methods replace zero-norm with a relaxed function and try to solve the new problem. Basis Pursuit [4], LASSO [5], Smoothed 
$L_0$ [6] and Iterative Reweighted Least Squares(IRLS)  [7] are instances of the second approach. 
Several sufficient conditions have been developed to formalize the notion of the suitability of a dictionary for sparse estimation. These include the mutual coherence [8], null space property [9], the Exact Recovery Coefficient [10], the spark [11] and the Restricted Isometry Constants (RICs) [12]. Except for the mutual coherence, none of these measures can be efficiently calculated for an arbitrary given dictionary \textbf{D}. For example, the RICs require enumerating over an exponential number of index sets in order to find the worst sub-dictionary. RICs consist of two constants: the Restricted Isometry Property (RIP) and the Restricted Orthogonality Property (ROP) constants. RIP is defined as follows:\\
\\
\textit{A dictionary is said to satisfy 
$RIP$
 of order
$\textbf{s}$
 with parameter 
$\delta_s$
if for every s-sparse 
$\textbf{\textit{x}}$ (i.e. 
$\textbf{\textit{x}}$ 
contains at most 
$\textbf{s}$
nonzero entries) the following inequality is consistent:\\
\begin{equation}
(1-\delta_s)\Vert \textbf{\textit{x}}\Vert_2^2\le \Vert \textbf{D}\textbf{\textit{x}}\Vert_2^2\le (1+\delta_s)\Vert \textbf{\textit{x}}\Vert_2^2
\end{equation}
}
According to 
$RIP$,
 the worst sub-dictionary (which bounds the inequality) corresponds to 
$\textbf{s}$ columns of 
$\textbf{D}$
 which have the minimum eigenvalue:\\
\\
\begin{equation}
\underset{\Omega}{\min} ~ \textit{eig}(\textbf{D}_{\Omega}^T \textbf{D}_{\Omega})\le \frac{\Vert \textbf{D}\textbf{\textit{x}}\Vert_2^2}{\Vert \textbf{\textit{x}}\Vert_2^2}\le \underset{\Omega}{\max} ~ \textit{eig}(\textbf{D}_{\Omega}^T \textbf{D}_{\Omega})
\end{equation}
Where
$\Omega$
 is the set of all possible atom combinations of 
$\textbf{D}$
. It’s obvious that the inequality of RIP is limited by columns that are linearly dependent. This paper aims at determining ill sub-dictionaries(i.e. those which have a small eigenvalue) to treat them in a special way. Combination of
$ l_1$
 minimization and full search is our motto which is guaranteed to recover the underlying coefficients under certain conditions. We will study these conditions.\\
In comparison to spark and dictionary coherency, RIP is a general criterion. By choosing 
$s=2$, the mutual coherence can be obtained by 
$\delta_2$:\\
\begin{equation}
\mu ~ = ~ \underset{i\ne j}{max}~ d_i^T d_j = 1 - \underset{\Omega}{\min} ~ \textit{eig}(\textbf{D}_{\Omega}^T \textbf{D}_{\Omega}) ~ = ~ \delta_2
\end{equation}
$\Omega$ is the set of ${K \choose 2}$ subsets of the dictionary and $\mu$ is the coherency of $\textbf{D}$. On the other hand, by putting $\delta_s = 1$, i.e., the minimum eigenvalue is fixed to zero or $\Vert \textbf{D}_{\Omega}\textbf{\textit{x}} \Vert_2^2 = 0$ the minimum corresponding $\textbf{s}$ is equal to spark.
\begin{equation}
spark = \underset{s}{\min}~ s ~~ \textit{s.t.} ~~ \delta_s = 1
\end{equation} 
These criterions provide us with few tools in order to ensure accurate recovery using different sparse coding algorithms [8-12]. \\
This paper first suggests to combine$ l_1$ minimization $(P_1)$ with $P_0$ to utilize the robustness of $l_1$ in cases where a strict RIP criterion is not met. This is done by specifying  an ill sub-dictionary and treating it in a more adequate manner. Such an approach enjoys much less computational burden than full search, although it may not be applicable in all situations. Inspired by the notion of specifying an ill subdictionary, we also introduce a practical method which can be applied on common sparse recovery algorithms to improve their performance.\\
The paper is organized as follows: In section II a novel combinatory algorithm is introduced. Then in section III a graphical tool is introduced to help improve the performance of sparse recovery algorithms. Sections IV and V investigate the implementation of the proposed tool in IRLS algorithm and present experimental results.\\

\section{Combination of $P_1$ \& $P_0$}
Let divide $\textbf{D}$ to two parts $\textbf{D}_1$ and $\textbf{D}_2$ such that the concatenation of $\textbf{D}_1$ with any $\textbf{P}$ columns of $\textbf{D}_2$ satisfies $RIP$, i.e. 
$N_p = {|D_2| \choose P}$ sub-matrices have $RIP$ with
$\delta_{2s} < \sqrt{2}-1$. According to (4) this constraint is less strict than $RIP$ for all the columns of $\textbf{D}$.\\   
It is shown that $P_0$ is guaranteed by $P_1$ for dictionaries that staisfy $RIP$ with the constant $\delta_{2s} < \sqrt{2} -1$ [17]. According to this theorem, the unique sparse solution will be obtained by solving $P_1$, for $ \textbf{y} = \textbf{D}^{(p)}\textbf{x}^{(p)} ~~\forall~ p = 1,\ldots, N_p$  if:\\
$~~~~~~~~~~~~~~~~~~~~~~~~~~~~~~~~\Vert x_2 \Vert_0 \le P$\\

\begin{figure}[t!]
\centering
\includegraphics[width=0.4\textwidth]{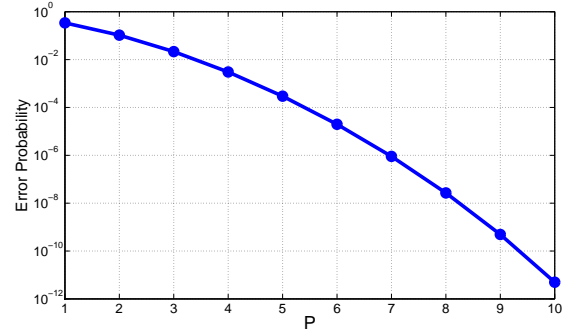}
\caption{\label{fig:12}Probability of error vs maximum number of non-zero elements in $\textbf{x}_2$
}
\end{figure}

\begin{proof}
consider all of the combinations of $\textbf{D}_2$ consist of $P$ columns. Concatenate $\textbf{D}_1$ with all those combinations to construct $N_P$ sets of columns. It is obvious that at least one of these sets include the exact support that we are looking for, if $\Vert \textbf{x}_2\Vert_0 \le P$ and by the assumption that all sets satisfy $RIP$ with the constant $\delta_{2s} < \sqrt{2} -1 $ which is a weaker condition, then $RIP$ holds for the set of all columns.
\end{proof}
In other words if the sparse solution contains at most $P$ non-zero coefficients corresponding to the second part, RIP guarantees to find the sparse solution after $N_P$ procedures of $l_1$ minimization, i.e. if there are more than $P$ non-zero coefficients in the second part, the sparse solution cannot be found. The probability of error can be written as the following form:\\
\begin{equation}
\footnotesize{P_e = \frac{\sum_{i=P+1}^{s}{|D_1| \choose s-i}{|D_2| \choose i}}{{K \choose s}} < (s-P)(\frac{s}{K-s+1})^{P-1}\frac{|D_2|^{P+1}s^3}{(P+1)!}\propto (\frac{s}{K})^P}
\end{equation}

For example, if $K=200$, $s=12$, $|\textbf{D}_2 |=20$ and $P=6$, the error would be $P_e=10^{-5}$ after solving ${20 \choose 6 }=38000$  "$P_1$"s which is far less than ${200 \choose 12 }=6.1\times 10^{18}$ for full search. Fig.1 shows probability of error versus parameter $P$ in the described setting.\\
The best choice for $\textbf{D}_2$ is the set of columns that make a small eigenvalue and keep RIP limited for $\textbf{D}$. Although computation load of this sparse recovery routine is much less than the full search, it is not practical in most of the cases but this approach brings us a good inspiration: Separation of ill columns and treating them in another way may result in an achievement for sparse recovery.\\
Now we present the results of a simulation that justifies the previous argument. Let assume an over-complete dictionary; we want to demonstrate that the number of sub-dictionaries that have a small eigenvalue and the error of sparse recovery are correlated. Let call a subset of the dictionary that has a small eigenvalue, an \textbf{ill sub-dictionary}. We actually want to show that coefficients corresponding to ill columns are more at risk for wrong recovery.\\
To performe the simulation, first assume a dictionary $\textbf{D}\in \mathbb{R}^{15\times25}$. Each atom alongside few other atoms may have a small eigenvalue. The ill sub-dictionaries are determined by combinational search among all possible 4 columns. We also define an error vector in which each entry shows the number of wrong estimations of a coefficient after running the algorithm for many times. For the assumed dictionary we performed $l_1$ minimization for sparse recovery. Sparsity is fixed to 7 to synthesize a signal and 200,000 different runnings are performed.\\

\begin{figure}[t!]
\centering
\includegraphics[width=0.4\textwidth]{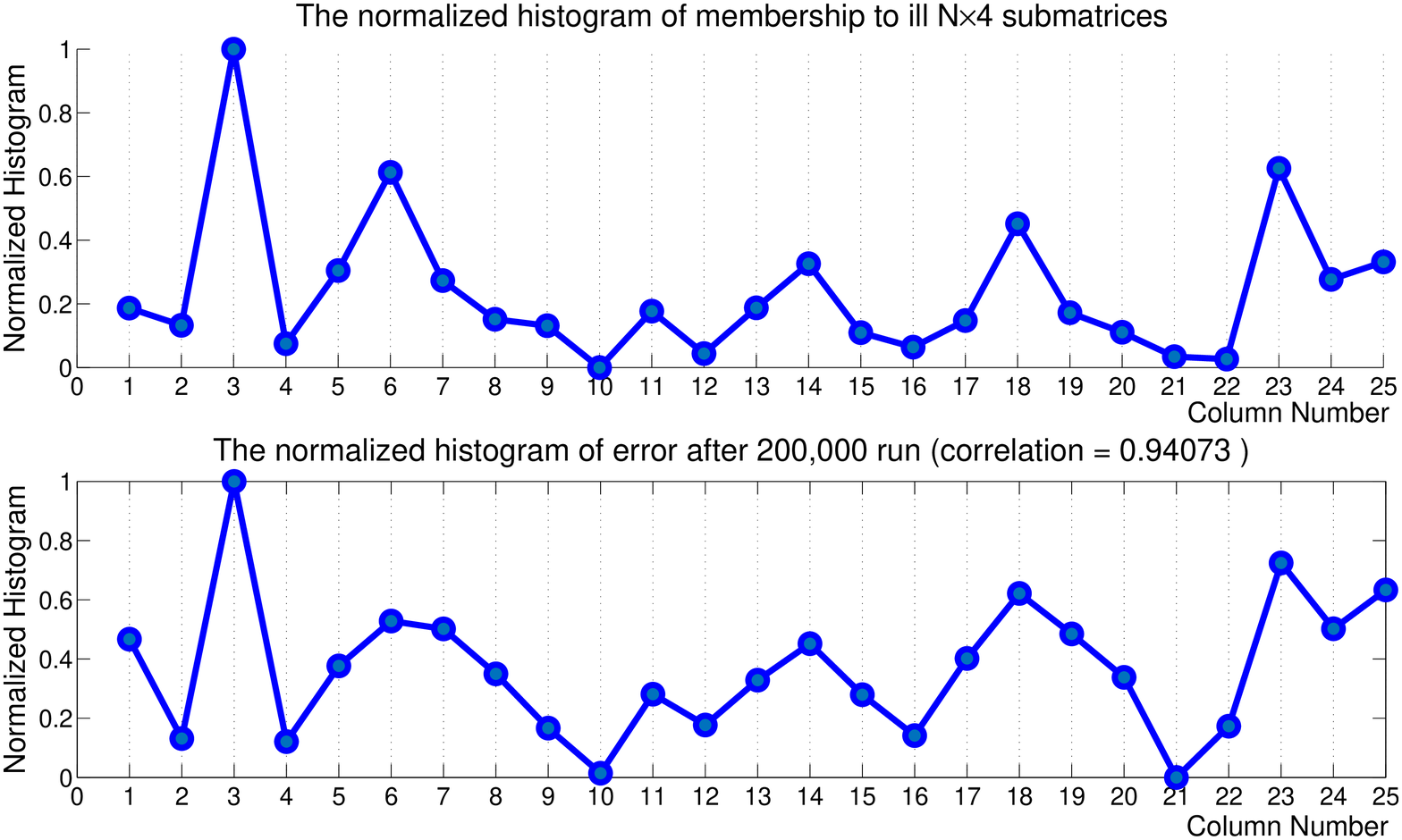}
\caption{\label{fig:12}Correlation between histogram of number of memberships in ill sub-dictionaries and sparse recovery error for 200,000 synthesized signals.
}
\end{figure}  
 Figure 2 plainly shows that the histogram of error(percentage of error per atom) and histogram of membership in ill sub-dictionaries are highly correlated. Table I shows the correlation when different search orders are used to obtain the histogram of membership in ill sub-dictionaries. The simulation confirms that more membership of a column in ill sub-dictionaries results in more error of sparse coding for that coefficient, which means that an ill sub-dictionary is able to mislead sparse recovery. In section 3 we introduce a new tool that successfully determines the atoms that may deceive a greedy algorithm in such a way.

\begin{table}[h]
\caption{Correlation between histogram of membership in ill sub-dictionary and histogram of error}
\begin{tabular}{ | l | p{1 cm} | p{1 cm} | p{1 cm} | p{1 cm} | p{1 cm} |}
    \hline
    Sparsity Order & 2 & 3 & 4&5&6 \\ \hline
    Correlation & 0.836 & 0.887 & 0.941 & 0.966 & 0.972 \\ \hline
    \end{tabular}\\
\end{table}
The question we are going to answer in the next section is "Is it possible to improve sparse recovery performance using prior information about the structure of the dictionary (like the dependence of error probability on the rate of memberships in ill subdictionaries as shown in Fig. 2)?"
\section{MATRIX COHERENCY GRAPH (MCG)}
Indication of ill sub-dictionaries is exploited in a practical and consistent manner in this section by suggesting to define a graph for a dictionary. To this aim we first determine sub-dictionaries that have small eigenvalues, then consider a graph with $K$ nodes and connect the nodes of an ill sub-dictionary by edges. The $\delta_s^{\Omega}$ in eq. (8) is the criterion by looking at which we construct edges of the graph corresponding to those $\Omega$ nodes.
\begin{equation}
\delta_s^{\Omega} = 1 - \min \{{\textit{eig}(\textbf{D}_{\Omega}^{T}\textbf{D}_{\Omega})}\}
\end{equation}
The graph is formed according to the following principle: If $\delta_s^{\Omega}\geq T$ then the nodes which correspond to $\Omega$ will be connected in the $MCG$ of $\textbf{D}$, where $T$ is a threshold between $0$ and $1$. Please note that minimization is driven on the eigenvalues of a certain subset of $\textbf{D}$ denoted by $\textbf{D}_{\Omega}$ to determine bad subsets; while in (4) the minimization is driven on $\Omega$ to determine the worst subset. Fig. 3 shows $MCG$ for the dictionary used in figure 2. Number of edges in this graph is proportional to values of figure 2 top. Columns connected together are linearly dependent - such that the mentioned threshold is surpassed - and can be used instead of each other. A greedy sparse recovery algorithm has to be more accurate in the case of connected columns, in order not to choose a wrong coefficient.

\begin{figure}[t!]
\centering
\includegraphics[width=0.4\textwidth]{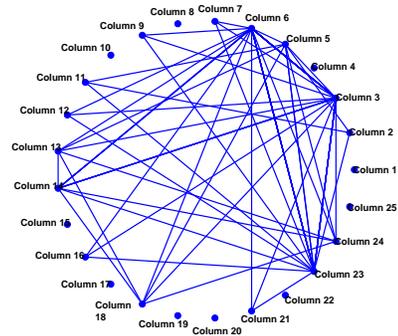}
\caption{\label{fig:12}An example of MCG for the dictionary used in Fig. 2.
}
\end{figure}

Order 3 $MCG$ with threshold $0.4$ for the over-complete $DCT$ [13] with redundancy factor $= 2$ in $\mathbb{R}^8$ is shown in figure 4.  In this case 16 bases are considered in $\mathbb{R}^8$ space.
Figure 5 shows an example of order 2 $MCG$ for the dictionary of $2D$ over-complete discrete cosine transform [13] in $\mathbb{R}^{(3\times3)}$. Redundancy factor of dictionary is 4, thus it has 36 atoms. Due to small eigenvalues of a set of connected columns, in the procedure of a greedy sparse coding algorithm they can be replaced by each other. As a result, to eliminate a column in a certain step of a greedy sparse coding algorithm, not only its coefficient should be small, but also its connected columns should correspond to small coefficients.\\
 Next section investigates using $MCG$ to improve $IRLS$ algorithm in the mentioned manner.

\section{USING MCG IN IRLS}
This paper introduced a new tool that can help sparse recovery algorithms to be less greedy. As the proposed tool is used to modify IRLS algorithm, we present a review on IRLS-based sparse recovery methods. IRLS, which is one of most accurate sparse coding algorithms, replaces the zero-norm by\\
\begin{equation}
f(\vec{x}) = \sum_{i=1}^{N}w_i x_i^2
\end{equation}
where, $x_i$ is the $i\textsuperscript{th}$ element of vector $\textbf{\textit{x}}$ and $w_i$ is selected such that the following equality is met\\
\begin{equation}
w_i = \frac{1}{|x_i^2| + \delta}
\end{equation}
In which $\delta$ is much smaller than the absolute value of the non-zero entries of $x$. The goal of the algorithm is solving the problem below:\\
\begin{equation} 
\underset{x}{\min}~x^T\textbf{W}x ~~ s.t. ~~ \textbf{\textit{y}} = \textbf{D}\textbf{\textit{x}}
\end{equation}	
Where $\textbf{W}= diag([w_1,\ldots, w_K ])$. As $\textbf{W}$ depends on $\textbf{\textit{x}}$ and is calculated in every sequential step, the above problem does not have a closed-form solution for $\textbf{\textit{x}}$. IRLS uses alternative optimization on $\textbf{\textit{x}}$  and $\textbf{W}$, where in each alternation, one of them is fixed and the other one is updated. The close-form solution for $\textbf{\textit{x}}$ in each alternation can be written as:\\

\begin{equation} \left \{
\begin{array}{lr}
\textbf{x}^l = \textbf{W}_l^{-1}\textbf{D}^T(\textbf{DW}_l^{-1}\textbf{D}^T)^{-1}\textbf{y}
\\
\\
\textbf{w}_i^l = \frac{1}{|\textbf{x}_i^{l-1}|^2 +\delta}
\end{array} \right. 
\end{equation} 

\begin{figure}[t!]
\centering
\includegraphics[width=0.5\textwidth]{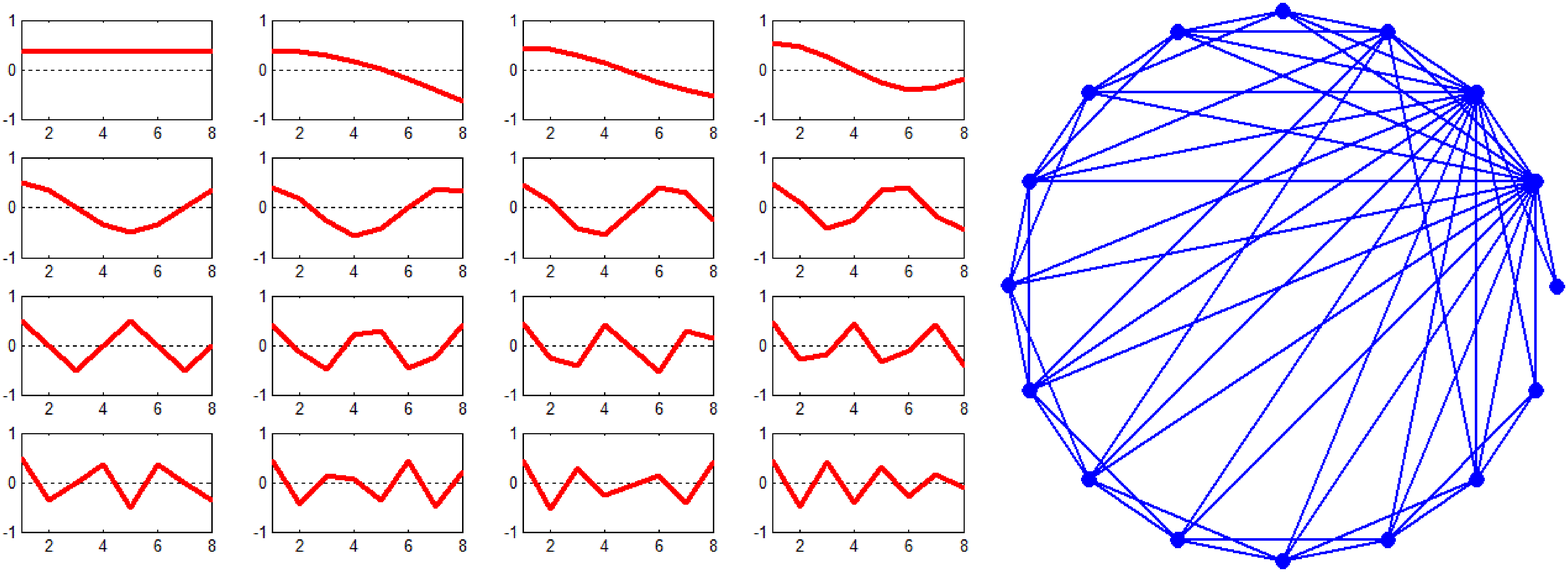}
\caption{\label{fig:12}Order 3 $MCG$ foran Overcomplete DCT dictionary with redundancy factor 2 in $\mathbb{R}^8$
}
\end{figure}

\begin{figure}[t!]
\centering
\includegraphics[width=0.45\textwidth]{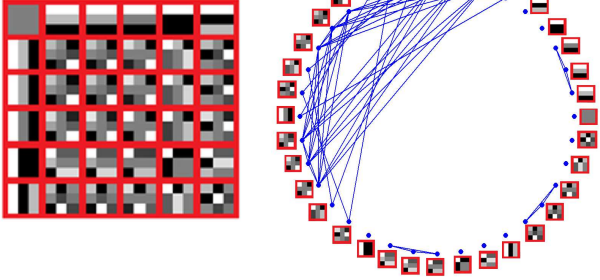}
\caption{\label{fig:12}Order 2 $MCG$ for 2D Overcomplete DCT with redundancy factor 4  in $\mathbb{R}^{3\times3}$
}
\end{figure}
Here, $\textbf{W}_l=diag([w_1^l,\ldots,w_K^l])$ and $l$ is the number of alternations. The algorithm needs initialization which is set as $\textbf{W}_0=I_{K\times K}$. If a coefficient is small in the current iteration, its penalty will be high in the next iteration and there will be no hope that it will survive as non-zero. Such a greedy approach seems to be the flaw of IRLS in many settings.\\
This algorithm can be used to address the problem of sparse recovery in the presence of noise,
\begin{equation}
\textbf{\textit{y}} = \textbf{D}\textbf{\textit{x}} + \textbf{\textit{n}}
\end{equation}
Where $n$ is white gaussian noise with variance $\sigma^2$. The closed-form solution is as follows:\\
\begin{equation}
\textbf{x} = (\textbf{D}^T\textbf{D}+\sigma^2\textbf{W})^{-1}\textbf{D}^T\textbf{y}
\end{equation}
As we mentioned in the previous section, those coefficients which are small as well as their connected coefficients can be eliminated. To apply this idea to $IRLS$, the weights should be updated as the following form:
\begin{equation}
w_i^l = (\frac{1}{|x_i^{l-1}|+\delta})^p(\frac{1}{f(x_{{\Omega}_i})})^q
\end{equation}
Where $f(x_{{\Omega}_i})$ is a function of connected coefficients to the $i\textsuperscript{th}$ coefficient ($\Omega_{i}$ is the set of indices including $i$ and indices of the atoms connected to the $i\textsuperscript{th}$  atom), for starting iterations put $q=2$ and $p=0$ and as the iterations go through we would have: $q\rightarrow0$ and $p\rightarrow2$ ($\textbf{x}^T\textbf{Wx}\rightarrow \Vert x\Vert_0$). A good choice for $f(x_{{\Omega}_i})$ is $\max(|x_{\Omega_{i}}|)$, by this choice connected coefficients to a large value coefficient still have the opportunity to contribute in the sparse code at the next iteration even if their values in the current iteration are small.
\section{EXPERIMENTAL RESULTS}
To evaluate the idea proposed in section II, consider the Gaussian random matrix $\textbf{D}\in \mathbb{R}^{15\times25}$ in figure 2 and 3. Let put the six columns have more ill-ness in $\textbf{D}_2$, i.e, high valued columns in figure 2 or high connected coulmns in figure 3. Figure 6 indicates sparse recovery versus cardinality of the solution. For each sparse recovery we have solved ${|\textbf{D}_2| \choose 4} = 15$ $l_1$ minimizations to increase the overall performance. In many practical situations it is not possible to use this approach. 
Fig. 7 shows the results of a simulation by Gaussian random matrix $\textbf{D}\in \mathbb{R}^{80\times160}$, which demonstrates successful recovery rate versus cardinality of the oracle solution. None of the parameters: number of non-zero entries, location of non-zero entries and value of them is known. In such order of dimensions, order 3 search is also practical.  
\begin{figure}[t!]
\centering
\includegraphics[width=0.45\textwidth]{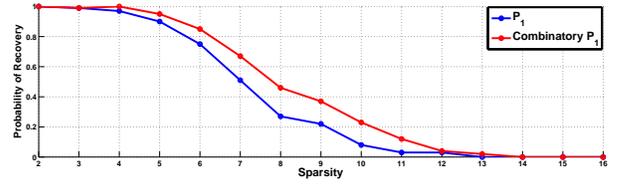}
\caption{\label{fig:12}Comparison of $P_1$ with combinatory $P_0 \& P_1$
}
\end{figure}

Error of the proposed sparse recovery is less correlated with comparison to what we have seen in table I. De-correlation means that the amount of error for coefficients corresponding to ill atoms is reduced. Table II compares the correlations.\\
Figure 8 shows number of iterations which is needed for algorithms to converge versus sparsity of the oracle solution.
\begin{figure}[h]
\centering
\includegraphics[width=0.45\textwidth]{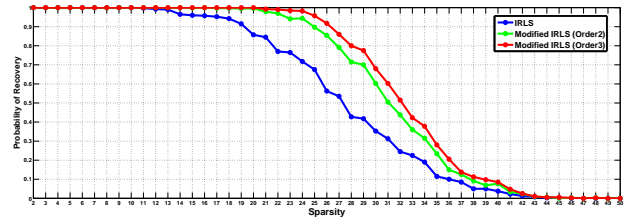}
\caption{\label{fig:12}Probability of recovery vs. Sparsity
}
\end{figure}

\begin{table}[h]
\caption{CORRELATION OF HISTOGRAM OF ILL SUB-DICTIONARY MEMBERSHIP AND HISTOGRAM OF ERROR }
\begin{tabular}{ | l | p{0.75 cm} | p{0.75 cm} | p{0.75 cm} | p{0.75 cm} | p{0.75 cm} |}
    \hline
    Search Order & 2 & 3 & 4&5&6 \\ \hline
    Correlation (Ordinary IRLS) & 0.81 & 0.88 & 0.94 & 0.95 & 0.96 \\ \hline
    Correlation (IRLS using MCG) & 0.69 & 0.75 & 0.83 & 0.84 & 0.85 \\ \hline
    \end{tabular}\\
\end{table}

\begin{figure}[h]
\centering
\includegraphics[width=0.45\textwidth]{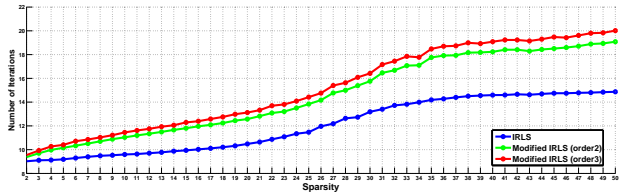}
\caption{\label{fig:12}Effect of applying $MCG$ on convergence rate of $IRLS$
}
\end{figure}

\section{CONCLUSION}
In this paper, we have presented a new approach for accurate sparse recovery. As we have shown, estimated coefficients by greedy algorithms have more error probability in the case of coefficients that their corresponding columns are members of an ill sub-dictionary. This prior information about probability of error is used to introduce Matrix Coherency Graph in our paper. We have exploited $MCG$ to modify $IRLS$ which is a well-known sparse recovery algorithm. The proposed tool also can be used in a way to tune parameters of other greedy sparse recovery algorithms and decrease their greediness.

\end{document}